\newif\ifShowKeys
\ShowKeysfalse

\documentclass[11pt]{article}
\pdfoutput=1
\usepackage[usenames,dvipsnames]{xcolor}
\usepackage[setpagesize=false,pagebackref=false, linktocpage, bookmarksopen=true, colorlinks=true, linkcolor=Maroon,citecolor=Maroon,urlcolor=Maroon]{hyperref}

\usepackage{tocloft}

\ifShowKeys \usepackage[notcite]{showkeys} \fi

\usepackage{amsmath, amssymb,amsthm}
\usepackage{amsthm}
\numberwithin{equation}{section}
\usepackage{array,longtable}

\usepackage{bm,environ,mathrsfs,array,arydshln}
\usepackage{booktabs,float,slashed,hyperref}
\usepackage[mathcal]{euscript}
\usepackage{tensor} 						
\usepackage{mathabx}
\usepackage{simpler-wick}
\usepackage[vcentermath]{youngtab}

\usepackage{aurical}
\usepackage[T1]{fontenc}
\usepackage[nodayofweek]{date time}
\usepackage{graphicx,epsfig,epic}
\usepackage{tikz} 
\usepackage{tikzfeynman}
\usetikzlibrary{arrows,decorations.pathreplacing,decorations.markings,snakes}
\tikzset{middlearrow/.style={decoration={markings, mark= at position 0.5 with {\arrow{#1}} ,
}, postaction={decorate}}}
\tikzset{decoration={snake,amplitude=.4mm,segment length=2mm,
                       post length=0mm,pre length=0mm}}

\usepackage{framed}						
\definecolor{shadecolor}{rgb}{0.9996078, 0.984314, 0.960784}

\usepackage{comment}

\allowdisplaybreaks

\definecolor{myred}{RGB}{233, 33, 45}

\newcommand{\bs}{\begin{shaded}}
\newcommand{\es}{\end{shaded}\noindent}
\def\ba#1\ea{\begin{align}#1\end{align}}		
\newcommand{\be}{\begin{equation}}
\newcommand{\ee}{\end{equation}}
\newcommand{\mc}{\mathcal }

\newcommand{\la}{\label}
\newcommand{\eps}{\varepsilon}

\newcommand{\lp}{\notag \\ & }

\newcommand{\cf}{\textit{cf.} }
\newcommand{\ie}{\textit{i.e.} }
\newcommand{\eg}{\textit{e.g.} }

\newcommand{\N}{\mathcal N}

\makeatletter
\DeclareFontFamily{OMX}{MnSymbolE}{}
\DeclareSymbolFont{MnLargeSymbols}{OMX}{MnSymbolE}{m}{n}
\SetSymbolFont{MnLargeSymbols}{bold}{OMX}{MnSymbolE}{b}{n}
\DeclareFontShape{OMX}{MnSymbolE}{m}{n}{
<-6>  MnSymbolE5
   <6-7>  MnSymbolE6
   <7-8>  MnSymbolE7
   <8-9>  MnSymbolE8
   <9-10> MnSymbolE9
  <10-12> MnSymbolE10
  <12->   MnSymbolE12
}{}
\DeclareFontShape{OMX}{MnSymbolE}{b}{n}{
<-6>  MnSymbolE-Bold5
   <6-7>  MnSymbolE-Bold6
   <7-8>  MnSymbolE-Bold7
   <8-9>  MnSymbolE-Bold8
   <9-10> MnSymbolE-Bold9
  <10-12> MnSymbolE-Bold10
  <12->   MnSymbolE-Bold12
}{}

\let\llangle\@undefined
\let\rrangle\@undefined
\DeclareMathDelimiter{\llangle}{\mathopen}%
 {MnLargeSymbols}{'164}{MnLargeSymbols}{'164}
\DeclareMathDelimiter{\rrangle}{\mathclose}%
 {MnLargeSymbols}{'171}{MnLargeSymbols}{'171}
\makeatother

\catcode`,\active

\catcode`\,12

\def\XXint#1#2#3{{\setbox0=\hbox{$#1{#2#3}{\int}$}
     \vcenter{\hbox{$#2#3$}}\kern-.5\wd0}}

\usepackage[object=vectorian]{pgfornament}

\newcommand{\I}{\mc I}

\newcommand{\s}{\sigma}

\DeclareMathOperator{\Pexp}{Pexp}
\DeclareMathOperator{\Res}{Res}
\newcommand{\res}[2]{\mathop{\Res}_{#1} #2}

\begin{document}


\begin{titlepage}

\vspace*{15mm}
\begin{center}
{\Large\sc   On the brane expansion of the Schur index  }\vskip 9pt

\vspace*{3mm}

{\large M. Beccaria${}^{\,a}$, A. Cabo-Bizet$^{\,a}$}

\vspace*{4mm}
	
${}^a$ Universit\`a del Salento, Dipartimento di Matematica e Fisica \textit{Ennio De Giorgi},\\ 
		and I.N.F.N. - sezione di Lecce, Via Arnesano, I-73100 Lecce, Italy
			\vskip 0.3cm
\vskip 0.2cm {\small E-mail: \texttt{matteo.beccaria@le.infn.it, acbizet@gmail.com}}
\vspace*{0.8cm}
\end{center}

\begin{abstract}  
\noindent
We consider the Schur index of $\N=4$ $U(N)$ SYM theory in 4d and its holographic giant graviton-type  expansion at finite $N$.
We compute the world-volume brane superconformal index by a recently proposed definition of the gauge holonomy integral
as a multivariate residue. This is evaluated by a novel deformation algorithm that avoids Gr\"obner basis methods. 
Various terms of the  brane expansion are computed and shown to be free of wall-crossing singularities to the order we explored. 
The relation between the brane expansion and previous giant graviton-type represenations of the Schur index is clarified. 
\end{abstract}
\vskip 0.5cm
	{
	}
\end{titlepage}

\tableofcontents
\vspace{1cm}

\section{Introduction and summary}

The superconformal index introduced in \cite{Romelsberger:2005eg,Kinney:2005ej,Bhattacharya:2008zy} 
may be regarded as the Witten index for superconformal theories in radial quantization. It is invariant under 
reasonable changes of model parameters and is therefore relevant to test weak-strong dualities.
Conversely, in the context of AdS/CFT, computing explicitly the index on both the gauge 
and string sides is an important test of the correspondence. This is non-trivial when one analyzes
the detailed dependence of the index on the state charges.
 
In the illustrative case of $\N=4$ $U(N)$ SYM on the world-volume of $N$ parallel D3 branes in type IIB superstring,
we may first  consider the index contributions from states with charges of order $N$. In this approximation, 
the index matches the counting of Kaluza-Klein (KK) BPS gravitons from supergravity on $AdS_{5}\times S^{5}$. 
For higher values of the charges, the index starts to depend on $N$, as it will be explained below. 
In this paper when we say large~$N$ expansion of the index we refer to its large~$N$ expansion at charges of order~$N$ or smaller.  
\footnote{This peculiar dependence on $N$
is important to reproduce black holes physics. Indeed, 
it is by now known that at large enough charges the asymptotic growth of the index is much faster than that of the gas of KK modes
\cite{Cabo-Bizet:2018ehj,Choi:2018hmj,Benini:2018ywd,Murthy:2022tbj,Agarwal:2020zwm}.}

Finite $N$ corrections to the index may be organized in additional series in
the index expansion parameter $q$ with overall weight $q^{nN}$, where $n=1, 2, \dots$ is the effect from $n$ giant gravitons 
\cite{McGreevy:2000cw,Grisaru:2000zn,Hashimoto:2000zp,Balasubramanian:2001nh,Mikhailov:2000ya}. 
This structure is  generic. For instance, in the S-fold background of IIB superstring \cite{Garcia-Etxebarria:2015wns}, beside giant gravitons,
one also has D3 branes wrapped on the internal space $S^{5}/\mathbb{Z}_{k}$, dual to Pfaffian-like operator, 
and these also provide additional finite $N$ corrections to the index \cite{Imamura:2016abe,Arai:2019xmp}. \footnote{The problem of counting states 
associated with these configurations was addressed in \cite{Biswas:2006tj,Mandal:2006tk,Kim:2006he}.}

In more details, in   $\N=4$ $U(N)$ SYM, finite $N$ corrections may be computed
by considering branes that are multiply wrapped around topologically trivial 3-cycles in the internal $S^{5}$ \cite{Imamura:2021ytr}.
Writing it as $|z_{1}|^{2}+|z_{2}|^{2}+|z_{3}|^{2}=1$, there are three 3-cycles defined by $z_{I}=0$. For wrapping
numbers $(n_{1},n_{2},n_{3})$ the gauge theory on the wrapped D3 branes one has a $U(n_{1})\times U(n_{2})\times U(n_{3})$ gauge theory
with bi-fundamental multiplets in a ring quiver diagram and the finite $N$ index is proposed to be 
\be
\la{1.1}
\I_{U(N)} = \I_{\rm KK}\, \sum_{n_{1},n_{2},n_{3}=0}^{\infty}\hat\I_{(n_{1},n_{2},n_{3})},
\ee
where $\I_{\rm KK}$  is the large $N$ Kaluza-Klein contribution.
The $N$ dependence of the brane index $\hat\I_{(n_{1},n_{2},n_{3})}$ is just from a classical prefactor coming from the 
classical charges and energy of the wrapped brane system of the schematic form $q^{(n_{1}+n_{2}+n_{3})N}$.
The actual calculation of the remaining part of $\hat\I_{(n_{1},n_{2},n_{3})}$ requires to integrate over the gauge holonomy
the plethystic exponential of an expression involving  the single-letter index of the brane world-volume superconformal theory. 

This step turns out to be subtle and rather non-trivial. First, the brane single-letter index includes tachyonic modes 
related to the topological triviality of the wrapping cycles. Their plethystic exponentiation requires an analytic continuation.
Second, the integration cycle of the gauge holonomy phases is not the naive one, where each belongs to the unit circle.
Instead, a prescription has to be given to determine which poles are kept and which are discarded. Although it is possible to 
match the gauge theory index at finite $N$ by a special choice of contour, it is unclear how to understand these rules in general.
In the analysis of  
\cite{Imamura:2021ytr}, wrapping up to $n_{1}+n_{2}+n_{3}=3$ was successfully considered, but extending the results to
higher windings remained a missing issue
due to the high algebraic complexity of the calculation.
Similar pole prescriptions appeared in other theories \cite{Arai:2019wgv,Arai:2019aou,Arai:2020uwd,Arai:2020qaj,Imamura:2021dya,Fujiwara:2023azx}
and a  discussion of what could be the correct pole prescription was presented in \cite{Imamura:2022aua}. \footnote{
In the recent paper \cite{Gautason:2023igo}, explicit string results for the worldsheet instantons in the ABJM theory \cite{Aharony:2008ug}
have been presented. In principle, they could be compared with the finite size corrections to the M2 brane index \cite{Arai:2020uwd} 
along the lines of \cite{Kim:2012ava}.
}

In \cite{Gaiotto:2021xce}, the problem was reconsidered trying to elucidate the precise analytic continuation relating the original 
gauge theory index and the brane index. The paradigmatic example is the finite $N$ half-BPS index of $\N=4$ $U(N)$ SYM 
that turned out to admit a simple expansion similar to (\ref{1.1})
\be
\la{1.2}
\I_{N}(q) = \I_{\infty}(q)\,[1+\sum_{k=1}^{\infty}q^{kN}\hat\I_{k}(q)], 
\ee
where the brane indices $\hat\I_{k}$ are given by the  remarkably simple analytic continuation rule \footnote{
We remark that expansions like (\ref{1.2}) for the superconformal index suffer from a certain ambiguity and is not unique. 
Indeed, in  \cite{Murthy:2022ien} it was given a general representation of that form for a class of matrix integrals over $U(N)$ that includes the superconformal index
integral representation. The $\hat\I$ functions in this construction are different from the ones arising from wrapped D3 branes
as pointed out in \cite{Liu:2022olj}. A critical discussion in the case of the half-BPS index appeared recently in  \cite{Eniceicu:2023uvd}.
}
\be
\la{1.3}
\hat \I_{k}(q) = \I_{k}(q^{-1}).
\ee
The holographic interpretation of (\ref{1.3})  is straightforward  \cite{Gaiotto:2021xce}. Let $X$ be an adjoint scalar 
and $q$ be the fugacity for a global symmetry $U(1)_{q}$ under which $X$ has unit charge.
Giant gravitons are D3 branes wrapped on the maximal $S^{3}$
fixed by $U(1)_{q}$ and have charge $N$. 
The corresponding radial fluctuations mode in the world-volume theory of the wrapped branes has $U(1)_{q}$ charge $-1$.
For a stack of $k$ giant gravitons,  the fluctuation mode is a $k\times k$ matrix of scalars and the world-volume theory has $U(k)$ gauge invariance. This suggests
indeed that the half-BPS excitations of wrapped branes are counted by the $U(k)$ index $\I_{k}(q^{-1})$, \ie (\ref{1.3}).

Let us remark that a very 
 non-trivial feature of (\ref{1.3}) is the fact that the inversion $q\to q^{-1}$ implies a resummation of the contributions to the index $\I_{k}$ after which it is possible to re-expand in powers
of $q$ for counting purposes.

This strategy was exploited systematically in   \cite{Gaiotto:2021xce} by considering other BPS sectors and models. Generally speaking, stacks of $k$ giant gravitons are dual to 
operators $(\det X)^{k}$ and fluctuations of the stack of wrapped D3 branes with $U(k)$ gauge theory on their world-volume are matched to finite modifications of determinant operators 
following \cite{Berenstein:2002ke,Balasubramanian:2002sa}.

Although the determinant modification strategy leads to a well-defined  single-letter index on the wrapped brane world-volume, the evaluation of the 
index still requires a crucial ingredient, \ie  again a prescription for the integration cycle of the gauge holonomy integral. This problem was addressed in full generality in 
\cite{Lee:2022vig} where a precise definition of the gauge  integral as a multivariate residue was proposed and  successfully tested in several examples.
\footnote{As remarked in  \cite{Gaiotto:2021xce},  relations like (\ref{1.2}) and its multi-fugacities generalizations are not simply combinatorial. Even if the superconformal gauge theory 
has not a weakly-curved holographic dual, one expects that in any $U(N)$ gauge theory is dual to a string theory where finite charge operators are 
dual to string excitations and operators of size $N$ can be associated to D-branes \cite{Witten:1979kh,Witten:1998xy}.}

In this paper, we explore the proposal of \cite{Lee:2022vig} in the case of the Schur index \cite{Gadde:2011uv}. This specialization of the full index has also 
been discussed in  \cite{Gaiotto:2021xce} and has the advantage of being computable at finite $N$ with minor effort, thanks to the methods and exact results
of  \cite{Bourdier:2015wda}. On the other hand, it is worth to revisit its brane expansion for various reasons.

We recall that the Schur index depends on two fugacities $x,y$ with a $\mathbb Z_{2}$
symmetry exchanging $x\leftrightarrow y$. In   \cite{Gaiotto:2021xce}, the index has been studied as a series in the parameter $q=xy$, followed by small $x$ expansion. 
The corresponding giant graviton-type representation was found to take the form
\be
\la{1.4}
\I_{N}(x; q) = \I_{\infty}(x; q)\sum_{k=0}^{\infty}x^{kN}\hat\I_{k}(x; q),
\ee
with the brane indices given by the specific analytic continuation
\be
\la{x15}
\hat\I_{k}(x; q) = \I_{k}(x^{-1}; x^{-1}q),
\ee
analogous to (\ref{1.3}).
On the other hand, as also suggested in \cite{Gaiotto:2021xce}, it should be possible to treat $x,y$ symmetrically and obtain a different kind of expansion \footnote{
We will denote by $F(x; q)$ quantities in terms of $x$ and $q=xy$, while $F(x,y)$ will be the same quantity in terms of $x,y$. So $F(x; xy) = F(x,y)$, but 
reason for this notation is that $F(x; q)$ will always be assumed to have been computed as a series expansion in powers of $q$, followed by expansion in $x$, 
while $F(x,y)$ will be later considered 
without a specific order of expansion.
}
\be
\la{1.6}
\I_{N}(x,y) = \I_{\infty}(x,y)\sum_{k,k'}x^{kN}y^{k'N}\hat\I_{(k,k')}(x,y).
\ee
This second representation is somehow more natural from the point of view of the wrapped D3 branes interpretation. \footnote{An expansion of the type (\ref{1.6}) was proposed
 in \cite{Arai:2020qaj} but their analysis needed a specific {\em ad hoc} pole prescription rule whose origin remained unclear.}

One additional reason to study the brane expansion of the symmetric Schur index is that 
looking at the small $x,y$ limit with fixed ratio $x/y$, one finds that the individual brane indices $\hat\I_{(k,k')}$ have rational terms with denominators having 
factors like powers of $x\pm y$. Such terms cannot be expanded unambiguously into a power series in $x$ and $y$. They are associated with  the wall-crossing 
phenomena discussed in \cite{Lee:2022vig} occurring when different fugacities or their powers collapse. These singularities should cancel since they are 
absent in the left hand side of (\ref{1.6}) \ie in the index of the original superconformal gauge theory. 

At the walls,  pole cancellation occurs and entails a peculiar enhancement of the brane index coefficients, \ie state degeneracy, that get an $N$ dependence. 
In  \cite{Lee:2022vig} it was suggested that in more physical cases, like the $\frac{1}{16}$-BPS index, 
this mechanism could be important to understand how  bulk microstates  build
emerging non-trivial geometries, \eg BPS black-holes.

\subsection*{Results}

To clarify the above issues, we computed various function $\hat\I_{(k,k')}$ in (\ref{1.6}) by the algorithm of \cite{Lee:2022vig}.
We did this by scaling
\be
(x,y)\to \eps\,(x,y),
\ee
where $\eps$ is a formal expansion parameter. This allows to analyze the regime of small $x,y$ with any fixed ratio $x/y$. The index $\I_{N}$ and 
the brane quantities $\hat\I_{(k,k')}$ are shown to admit a regular $\eps$-expansion. Remarkably, the leading contribution is always a non-trivial rational function of $x,y$. For example
\be
\hat\I_{(1,0)}(\eps x,\eps y) = -\frac{x^{2}}{x-y}\eps+\mc O(\eps^{2}), \qquad
\hat\I_{(2,0)}(\eps x,\eps y) = -\frac{x^{7}(x-2y)}{(x-y)^{2}y(x+y)}\eps^{4}+\mc O(\eps^{5}),
\ee
and so on. The presence of such contributions signals an ambiguity related to the order of the double expansion in $x$, $y$. In the language of 
\cite{Lee:2022vig} this is a wall-crossing phenomenon in the sense that two different expansions should be used 
depending on $|x/y|$ being smaller or larger than 1. 
This splits the fugacity space into two regions separated by the codimension-1 wall $x=y$.

In the separate  terms $\hat\I_{(k,k')}$, the limit $x\to y$ gives rise to a true singularity. Nevertheless, it has to cancel in the full finite $N$ index 
because it has a regular polynomial
dependence on $x,y$ order by order in $\eps$,
\be
\I_{N}(\eps x,\eps y) = 1+\sum_{n=1}^{\infty} \mc P_{N}^{(n)}(x,y)\,\eps^{n},
\ee
where $\mc P_{N}^{(n)}(x,y)$ are symmetric polynomials of degree $n$. For the quantities we have computed, we checked that this cancellation 
indeed occurs by considering the subset of terms with fixed level $k+k'$.

Our explicit multivariate residue calculation also clarifies the relation between the two apparently incompatible expansions (\ref{1.4}) and (\ref{1.6}). 
We will argue that 
\be
\la{1.10}
\hat\I_{k}(x; q) \equiv \I_{k}(x^{-1}; y) = \hat\I_{(k,0)}(x,y).
\ee
In this relation the l.h.s. is obtained by an analytic continuation of the gauge theory index 
according to (\ref{x15}), while the r.h.s. is the result from the multivariate residue computation.
Also, it is remarkable that the terms in (\ref{1.6}) that are missing in (\ref{1.4}), \ie those with with $k'>0$,  
do not contribute if (\ref{1.6}) is evaluated by expanding first in small $y$ and then in $x$, \ie in the 
asymmetric limit where (\ref{1.4}) is known to hold. This is the only regime where (\ref{1.4}) and (\ref{1.6}) are equivalent, while
in the more general case of fixed ratio $x/y$, the correct expansion is necessarily the  double sum in  (\ref{1.6}).

We also examined the structure of the index expansion at the wall $x=y$ and could confirm the peculiar enhancement
of its coefficients that are  polynomials in $N$. In more details, the index takes the form 
\ba
 \la{1.11}
 \I_{N}(x, y)\bigg|_{x=y} =& \I_{\infty}(x)\, \bigg[1+\sum_{k=1}^{\infty}Q_{k}(N)x^{kN+k^{2}}\bigg],
 \ea
where $\I_{\infty}$ has a regular series in $x$ independent on $N$ and $Q_{k}(N)$ are computable polynomials in $N$ of degree $k$.
Relation (\ref{1.11}) follows from the known expression of the index at the wall \cite{Bourdier:2015wda}. In the brane expansion, 
it is a consequence of the wall-crossing poles cancellation.

\bigskip

The plan of the paper is the following. In section \ref{sec:GL} we summarize the Gaiotto-Lee construction of the brane single-letter index
from determinant operator modifications as the prescription to define the brane (full multi-particle) indices as multivariate residue.
In section \ref{sec:schur} we discuss the Schur index in the $\eps$-expansion. We compute various brane indices as multivariate residue by 
a novel deformation algorithm. Cancellation of wall-crossing poles is checked up to level $k+k'=3$ and we verify the validity of the double expansion (\ref{1.6}).
In section \ref{sec:comp}
we compare the two expansions (\ref{1.4}) and (\ref{1.6}) showing how resummation of the $q=xy$ is possible to achieve the remarkable equality (\ref{1.10}).
Finally, in section \ref{sec:deg} we discuss the degeneracy enhancement summarized in (\ref{1.11}) and happening at the wall $x=y$.
In appendix \ref{app:quart} we briefly discuss the case of the solvable $\frac{1}{4}$-BPS Schur index where similar mechanisms are illustrated.

\section{Gaiotto-Lee determinant modification construction}
\la{sec:GL}

Let us briefly summarize the Gaiotto-Lee determinant modification construction to 
build the single letter brane index and the prescription given in \cite{Lee:2022vig}
 to compute the 
brane index by integrating out gauge fugacities.
We consider a $U(N)$ supersymmetric gauge theory and represent the finite $N$ index $\I_{N}(x)$
depending on a set of counting variables $\bm{x}=(x_{1}, \dots, x_{s})$ in the form of a giant graviton-type expansion
\be
\la{2.1}
\I_{N}(\bm{x}) = \I_{\infty}(\bm{x})\sum_{k_{1}, k_{2}, \dots, k_{s}=0}^{\infty}x_{1}^{k_{1}N}\cdots x_{s}^{k_{s}N}\, \hat{\I}_{(k_{1}, \dots, k_{s})}(\bm{x}).
\ee
Here, $\I_{\infty}(\bm{x})$ is the large $N$ index equal to the 
(dual) closed string index given by the fluctuations of Kaluza-Klein supergravity modes. The sum in (\ref{2.1})
is over the wrapping number of branes wrapped on different supersymmetric 
cycles. Each term in (\ref{2.1}) is associated with a stack of $k_{i}$ branes of type $i$ and the brane index $\hat\I_{(k_{1}, \dots, k_{s})}$ 
counts states in the worldvolume $\prod_{i}U(k_{i})$ quiver gauge theory. In the dictionary of \cite{Lee:2022vig}, these states are referred to  as
open string excitations on the $(k_{1}, \dots, k_{s})$ brane stack.

Adjoint fields in the $U(N)$ gauge theory are counted by the single-letter index $f(\bm{x})$ which is a rational function of the 
fugacities $\bm{x}$ associated with the global symmetries of the gauge theory.
The total (multi-letter) index of the gauge theory is written in the usual way as an integral over $U(N)$ gauge fugacities -- with 
standard integration cycle -- 
\be
\I_{N}(\bm{x}) = \frac{1}{N!}\oint_{|z_{a}|=1}\prod_{a=1}^{N}\frac{dz_{a}}{2\pi i z_{a}}\prod_{a\neq b}^{N}\bigg(1-\frac{z_{a}}{z_{b}}\bigg)\,
\Pexp\bigg[f(\bm{x})\,\sum_{a,b=1}^{N}\frac{z_{a}}{z_{b}}\bigg],
\ee
where $\Pexp$ denotes the plethystic exponential.
At large $N$, one has simply
\be
\I_{\infty}(\bm{x})  = \prod_{n=1}^{\infty}\frac{1}{1-f(\bm{x}^{n})}, \qquad \bm{x}^{n} = (x_{1}^{n}, \dots, x_{s}^{n}).
\ee
The prescription in \cite{Gaiotto:2021xce} counts modifications of the determinant product
$\prod_{i=1}^{s}(\det X_{i})^{k_{i}}$,
where $X_{i}$ are the adjoint fields, and modification means that we can replace in $\det X_{i}$ the letter $X_{i}$ somewhere by other fields.
As proven in \cite{Gaiotto:2021xce,Lee:2022vig}, this provides the representation (\ref{2.1}) with the following expression for the brane index
\ba
\la{2.4}
\hat\I_{(k_{1}, \dots, k_{s})}(\bm{x}) &= \frac{1}{k_{1}!\cdots k_{s}|!}\oint\prod_{a_{1}=1}^{k_{1}}\frac{d\s^{X_{1}}_{a_{1}}}{2\pi i \s^{X_{1}}_{a_{1}}}\cdots
\prod_{a_{s}=1}^{k_{s}}\frac{d\s^{X_{s}}_{a_{s}}}{2\pi i \s^{X_{s}}_{a_{s}}}\lp
\prod_{a_{1}\neq b_{1}}\bigg(1-\frac{\s^{X_{1}}_{a_{1}}}{\s^{X_{1}}_{b_{1}}}\bigg)\cdots
\prod_{a_{s}\neq b_{s}}\bigg(1-\frac{\s^{X_{s}}_{a_{s}}}{\s^{X_{s}}_{b_{s}}}\bigg)\, \Pexp\bigg[\sum_{i,j=1}^{s}\hat f^{i}_{j}\sum_{a_{i},b_{j}}\frac{\s^{X_{i}}_{a_{i}}}{\s^{X_{j}}_{b_{j}}}\bigg],
\ea
where the modified single-letter index is 
\be
\la{2.5}
\hat f^{i}_{j}(\bm{x}) = \delta^{i}_{j}+\frac{(x_{i}-1)(1-x_{j}^{-1})}{1-f(\bm{x})},
\ee
and the integration cycle will be discussed in section \ref{sec:cycle}.
As an example, in the half-BPS sector we have a single letter $X$ and the determinant operator
$\det X = \frac{1}{N!}\eps^{i_{1}\cdots i_{N}}\eps_{j_{1}\cdots j_{N}}X^{j_{1}}_{i_{1}}\cdots X^{j_{N}}_{i_{N}}$
has weight $x^{N}$. The only admissible modification is $X\to 1$ that reduce by 1 the exponent of $x$. 
This is consistent with (\ref{2.5}) that gives using $f = x$
\be
\hat f (x)= 1+\frac{(x-1)(1-x^{-1})}{1-x} = x^{-1}.
\ee
From the general formula  (\ref{2.4}),   modifications of $(\det X)^{k}$ in the half-BPS sector are described by the brane index
\ba
\la{2.7}
\hat\I_k(x) &= \frac{1}{k!}\oint\prod_{a=1}^{k}\frac{d\s_{a}}{2\pi i \s_{a}}
\prod_{a\neq b}\bigg(1-\frac{\s_{a}}{\s_{b}}\bigg)\Pexp\bigg[\frac{1}{x}\sum_{a,b}\frac{\s_{a}}{\s_{b}}\bigg],
\ea
The expected result is the explicit formula,\cf (\ref{1.3}), 
\be
\la{2.8}
\hat \I_{k}(x) = \I_{k}(x^{-1}) = (-1)^{k}\frac{x^{k(k+1)/2}}{\prod_{m=1}^{k}(1-x^{m})},
\ee
where we used $\I_{N}(x) = 1/\prod_{n=1}^{N}(1-x^{n})$.

\subsection{Analytic continuation}
\la{sec:cycle}

As we mentioned, the expression (\ref{2.4}) still requires to  specify how to integrate over the parameters $\s_{a}^{X_{a}}$.
As in the example of the half-BPS sector,  the world-volume theory contains field with opposite charges to the ones of the original theory.
In the half-BPS sector, one can compute the index $\hat\I_{k}$ for modifications of $(\det X)^{k}$ as a power series in $x^{-1}$ and then analytically 
continue its resummation to a power series in $x$. In general, the relation between $\hat f$ in (\ref{2.5}) and $f$ is not so simple and one needs an independent way
to evaluate (\ref{2.4}). The proposal in \cite{Lee:2022vig} is to compute it as a multivariate residue with a rather general prescription of the 
canonical integration cycle \cite{Larsen:2017aqb}. 

We recall that multivariate residues occur in our context for an integrand of the form 
\be
\la{2.9}
\frac{h(\bm\s)\,d\s_{1}\wedge\cdots d\s_{K}}{g_{1}(\bm\s)\cdots g_{K}(\bm\s)}, \qquad K=\sum_{i}k_{i},
\ee
with the point $\bm\s=0$ being an isolated common zero of the denominator factors $g_{a}(\bm\s)$. 
The canonical integration cycle is the torus $|g_{a}(\bm\s)|=\eps$ for small enough $\eps$, and 
orientation $d(\text{arg}\, g_{1})\wedge\cdots\wedge d(\text{arg}\, g_{K})\ge 0$.
Unlike the 1-dimensional case, it  depends on the detailed factors $g_{a}(\bm\s)$ in the denominator of (\ref{2.9})
and not just on the full denominator. Notice also that the integration cycle is not the trivial one $|\s_{a}|=\eps$.

To define the factors $g_{a}(\bm \s)$, we recall that the integrand of $\hat\I_{(k_{1}, \dots, k_{s})}$ generally involves ratios of infinite products.
Numerators and denominators come respectively from negative and positive terms that
appear in an expansion of the brane single-letter index $\hat f^{i}_{j}(\bm x)$. If for some $i,j$ we have the expansion in monomials
$\hat f^{i}_{j}(\bm x) = +\sum_{\alpha}p_{\alpha}(\bm x)-\sum_{\beta}n_{\beta}(\bm x)$, 
we will find  in the denominator of the integrand products of factors
\be
\s^{X_{i}}_{a_{i}}-p_{\alpha}(x)\s^{X_{j}}_{b_{j}}.
\ee
Interpretation of this factor is a single open string connecting $(\det X_{j})_{b_{j}}$ with $(\det X_{i})_{a_{i}}$. 
Following  \cite{Lee:2022vig}, we will partition the denominator as 
\be
\la{2.11}
g = \{(g^{X_{1}}_{1}, \dots, g_{k_{1}}^{X_{1}}), (g^{X_{2}}_{1}, \dots, g_{k_{2}}^{X_{2}}),\cdots 
(g^{X_{s}}_{1}, \dots, g_{k_{s}}^{X_{s}})\}. 
\ee
In this partition
we put in $g^{X_{1}}_{1}$ all denominators that represent open strings ending on $(\det X_{1})_{1}$. Then we put in $g^{X_{1}}_{2}$
all remaining denominators that represent open strings ending on $(\det X_{1})_{2}$ and so on. If in this ordering $g^{X_{i}}_{a_{i}}$ comes before $g^{X_{j}}_{b_{j}}$ 
then we place into 
$g^{X_{i}}_{a_{i}}$ the factor 
\be
\s^{X_{i}}_{a_{i}}-p_{\alpha}(x)\s^{X_{j}}_{b_{j}} \to \s^{X_{j}}_{b_{j}}-p_{\alpha}(x)^{-1}\s^{X_{i}}_{a_{i}}.
\ee

\subsection{A deformation algorithm for the multivariate residue}

To illustrate the computational difficulties in the evaluation of the multivariate residue associated with the cycle (\ref{2.11}), let us consider again
the half-BPS case. From  (\ref{2.7}), evaluating the plethystic, we have  
\ba
\la{2.13}
\hat\I_k(x) &
= \frac{1}{k!}\bigg(\frac{1}{1-x^{-1}}\bigg)^{k}\,\oint\prod_{a=1}^{k}\frac{d\s_{a}}{2\pi i \s_{a}}
\prod_{a\neq b}\frac{\s_{a}-\s_{b}}{\s_{a}-x^{-1}\s_{b}},
\ea
and the proposed partitioning of the denominator is
\be
g = \{g_{1}, \dots, g_{k}\}, \qquad g_{a} = \s_{a}\, \prod_{b\neq a}(\s_{a}-x^{-1}\s_{b}).
\ee
To compute the multivariate residue it is important to determine whether the pole is non-degenerate, \ie has non-vanishing Jacobian
\be
J = \det_{a,b}\left(\frac{\partial g_{a}}{\partial\sigma_{b}}\right)\bigg|_{\sigma=0}.
\ee
If $J\neq 0$ at the pole $\s=\s^{*}$, then we have simply
\be
\la{2.16}
\res{\s=\s^{*}; g}{\omega} = \frac{h(\s^{*})}{J(\s^{*})}.
\ee
If the pole is degenerate ($J=0$) there are algebraic geometry algorithms to compute the multivariate residue, based on Gr\"obner basis methods. These 
become rapidly useless in our cases due to (i) the presence of the fugacities as free parameters, and (ii) the fact that the plethystic exponential produces
(generically although not in half-BPS case) infinite products that have to be truncated to a large number of factors to get accurate results for series expansions of the brane indices.

We have used a trick based on a deformation of the integrand in order to deal with non-degenerate poles only. 
In the half-BPS case, the integral (\ref{2.13}) reproduces (\ref{2.8}) if  the function
\be
G_{k}(x) = \oint\prod_{a=1}^{k}\frac{d\s_{a}}{2\pi i \s_{a}}
\prod_{a\neq b}^{k}\frac{\s_{a}-\s_{b}}{\s_{a}-x^{-1}\s_{b}},
\ee
has the expression
\be
\la{218}
G_{k}(x) = k!\,x^{k(k-1)/2}\,\prod_{m=1}^{k}\frac{1-x}{1-x^{m}},
\ee
which is what we want to obtain. We deform the integrand by considering
\ba
\la{2.19}
G_{k}^{(\eps)} (x)
=\oint\prod_{a=1}^{k}\frac{d\s_{a}}{2\pi i (\s_{a}+\eps_{a})}
\prod_{a\neq b}\frac{\s_{a}-\s_{b}}{\s_{a}-x^{-1}\s_{b}+\eps_{ab}},
\ea
where, in this simple case, we just choose
\be
\eps_{a} = a\,\eps,\qquad \eps_{ab}=\kappa\eps,
\ee
where $\eps$, $\kappa$ are parameters that for the moment are not fixed. Notice however that it is important to have different $\eps_{a}$
for different $a$. The poles are now all non-degenerate.
Since the deformation is a simple shift, the computation of the set of poles is not demanding. For higher $k$ the number of poles $p_{k}$ is 
\be
p_{2}=4, \ p_{3}=20, \ p_{4} = 136, \ p_{5}=1182, \ p_{6}=12304, \cdots.
\ee
This number grows quickly but slow enough to allow to go well beyond what can be reached with the standard methods.

 For example, for $k=2$, the pole $\s_{a}=0$ splits into the following four poles in $(\s_{1},\s_{2})$ whose residue 
 may be computed by (\ref{2.16})
\be
\nopagebreak
\def\arraystretch{1.3}
\begin{array}{lcc}
\toprule
& \textsc{pole} & \textsc{residue}  \\
\midrule
1&(-\eps , -2 \eps) & -\frac{x^2}{(1-2 x+\kappa x) (2-x+\kappa x)}  \\
2&( -\eps,  -\kappa \eps -\frac{\eps }{x}) & \frac{x (1-x+\kappa x)}{(1+x) (1-2 x+\kappa x)} \\
3&( -\frac{\kappa x \eps }{-1+x} , -\frac{\kappa x \eps }{-1+x}) & 0  \\
4&( -\kappa \eps -\frac{2 \eps }{x} , -2 \eps)  &\frac{x (2-2 x+\kappa x)}{(1+x) (2-x+\kappa x)}  \\
\bottomrule
\end{array}
\ee
The residue is independent on $\eps$ showing that the deformed cycle is in the same class, \ie $\hat\I_k(x; \eps)$ in (\ref{2.19})
is actually independent on the deformation parameter $\eps$. The single residues depend still on $\kappa$, but the sum does not and 
reads
\be
G_{2}(x) = \sum\text{Residues} = \frac{2x}{1+x},
\ee
which is the correct value to reproduce $\hat \I_{2}(x)$. For the next value $k=3$ there are 20 poles $(\s_{1},\s_{2},\s_{3})$ and the associated residues are 
\be
\def\arraystretch{1.3}
\begin{array}{lcc}
\toprule
&\textsc{pole} & \textsc{residue}  \\
\midrule
1& (-\eps ,-2 \eps ,-3 \eps ) & -\frac{4 x^6}{(1-3 x+x \kappa 
) (2-3 x+x \kappa ) (1-2 x+x \kappa ) (3-2 x+x \kappa ) (2-x+x \kappa 
) (3-x+x \kappa )} \\
2& (-\eps ,-2 \eps ,-\frac{\eps +x \eps  \kappa }{x}) & 
\frac{x^4 (1-2 x+x \kappa ) (1-x+x \kappa )}{(1+x) (1-3 x+x \kappa ) 
(2-x+x \kappa ) (1-2 x^2+x \kappa +x^2 \kappa )} \\
 3&(-\eps ,-2 \eps ,-\frac{2 \eps +x \eps  \kappa }{x}) & 
-\frac{x^4 (2-2 x+x \kappa ) (2-x+x \kappa )}{(1+x) (2-3 x+x \kappa ) 
(1-2 x+x \kappa ) (2-x^2+x \kappa +x^2 \kappa )} \\
 4&(-\eps ,-\frac{x \eps  \kappa }{-1+x},-\frac{x \eps  
\kappa }{-1+x}) & 0 \\
 5&(-\eps ,-\frac{\eps +x \eps  \kappa }{x},-3 \eps ) & 
\frac{2 x^4 (1-3 x+x \kappa ) (1-x+x \kappa )}{(1+x) (1-2 x+x \kappa 
) (3-x+x \kappa ) (1-3 x^2+x \kappa +x^2 \kappa )} \\
 6&(-\eps ,-\frac{\eps +x \eps  \kappa }{x},-\frac{\eps +x 
\eps  \kappa }{x}) & 0 \\
 7&(-\eps ,-\frac{\eps +x \eps  \kappa }{x},-\frac{\eps +x 
\eps  \kappa +x^2 \eps  \kappa }{x^2}) & \frac{x^3 (1-x+x \kappa 
)^2}{(1+x+x^2) (1-2 x+x \kappa ) (1-3 x^2+x \kappa +x^2 \kappa )} \\
 8&(-\eps ,-\frac{3 \eps +x \eps  \kappa }{x},-3 \eps ) & 
-\frac{2 x^4 (3-3 x+x \kappa ) (3-x+x \kappa )}{(1+x) (1-3 x+x \kappa 
) (3-2 x+x \kappa ) (3-x^2+x \kappa +x^2 \kappa )} \\
 9&(-\eps ,-\frac{\eps +x \eps  \kappa +x^2 \eps  \kappa 
}{x^2},-\frac{\eps +x \eps  \kappa }{x}) & \frac{x^3 (1-x+x \kappa 
)^2}{(1+x+x^2) (1-3 x+x \kappa ) (1-2 x^2+x \kappa +x^2 \kappa )} \\
 10&(-\frac{x \eps  \kappa }{-1+x},-2 \eps ,-\frac{x \eps  
\kappa }{-1+x}) & 0 \\
\bottomrule
\end{array}\notag
\ee
and 
\be
\def\arraystretch{1.3}
\begin{array}{lcc}
\toprule
&\textsc{pole} & \textsc{residue}  \\
\midrule
11& (-\frac{x \eps  \kappa }{-1+x},-\frac{x \eps  \kappa 
}{-1+x},-3 \eps ) & 0 \\
 12&(-\frac{x \eps  \kappa }{-1+x},-\frac{x \eps  \kappa 
}{-1+x},-\frac{x \eps  \kappa }{-1+x}) & 0 \\
 13&(-\frac{2 \eps +x \eps  \kappa }{x},-2 \eps ,-3 \eps ) & 
\frac{x^4 (2-3 x+x \kappa ) (2-2 x+x \kappa )}{(1+x) (3-2 x+x \kappa 
) (2-x+x \kappa ) (2-3 x^2+x \kappa +x^2 \kappa )} \\
 14&(-\frac{2 \eps +x \eps  \kappa }{x},-2 \eps ,-\frac{2 \eps 
+x \eps  \kappa }{x}) & 0 \\
 15&(-\frac{2 \eps +x \eps  \kappa }{x},-2 \eps ,-\frac{2 \eps 
+x \eps  \kappa +x^2 \eps  \kappa }{x^2}) & \frac{x^3 (2-2 x+x \kappa 
)^2}{(1+x+x^2) (2-x+x \kappa ) (2-3 x^2+x \kappa +x^2 \kappa )} \\
 16&(-\frac{3 \eps +x \eps  \kappa }{x},-2 \eps ,-3 \eps ) & -\
\frac{x^4 (3-3 x+x \kappa ) (3-2 x+x \kappa )}{(1+x) (2-3 x+x \kappa 
) (3-x+x \kappa ) (3-2 x^2+x \kappa +x^2 \kappa )} \\
 17&(-\frac{3 \eps +x \eps  \kappa }{x},-\frac{3 \eps +x \eps  
\kappa }{x},-3 \eps ) & 0 \\
 18&(-\frac{3 \eps +x \eps  \kappa }{x},-\frac{3 \eps +x \eps  
\kappa +x^2 \eps  \kappa }{x^2},-3 \eps ) & \frac{x^3 (3-3 x+x \kappa 
)^2}{(1+x+x^2) (3-x+x \kappa ) (3-2 x^2+x \kappa +x^2 \kappa )} \\
 19&(-\frac{2 \eps +x \eps  \kappa +x^2 \eps  \kappa }{x^2},-2 
\eps ,-\frac{2 \eps +x \eps  \kappa }{x}) & \frac{x^3 (2-2 x+x \kappa 
)^2}{(1+x+x^2) (2-3 x+x \kappa ) (2-x^2+x \kappa +x^2 \kappa )} \\
 20&(-\frac{3 \eps +x \eps  \kappa +x^2 \eps  \kappa \
}{x^2},-\frac{3 \eps +x \eps  \kappa }{x},-3 \eps ) & \frac{x^3 (3-3 
x+x \kappa )^2}{(1+x+x^2) (3-2 x+x \kappa ) (3-x^2+x \kappa +x^2 
\kappa )} \\
\bottomrule
\end{array}\notag
\ee
The sum of residues is again independent on $\kappa$ and in agreement with (\ref{218}).
We have checked that in this way the correct  $G_{k}(x)$ in (\ref{218})
are reproduced up to $k=6$
that is definitely out of reach even in this simple case if standard algorithms based on the transformation theorem \cite{cattani1996computing}
are used.

\section{Brane expansion of the Schur index}
\la{sec:schur}

The single-letter superconformal index for $\N=4$ $U(N)$ SYM depends on five fugacities with one constraint \cite{Kinney:2005ej}
\be
\la{3.1}
f(x,y,z,p,q) = 1-\frac{(1-x)(1-y)(1-z)}{(1-p)(1-q)}, \qquad xyz=pq.
\ee
The Schur index is the specialization $p=z$ \footnote{As in \cite{Gaiotto:2021xce} we consider the index in the Ramond sector to 
simplify the formulas. This is not a limitation since the Neveu-Schwarz index is obtained by the shift $x\to x\sqrt{q}$.} and its one-letter index is thus
\be
\la{3.2}
f = \frac{1}{1-q}\bigg(x+\frac{q}{x}-2q\bigg).
\ee
The finite $N$ index has expression \cite{Spiridonov:2010qv}
\ba
\la{3.3}
\I_{N}(x; q) =& \frac{1}{N!}\oint_{|\s_{a}|=1}\prod_{a=1}^{N}\frac{d\s_{a}}{2\pi i \s_{a}} \frac{\prod_{a<b}^{N}(1-\s_{a}\s_{b}^{-1})(1-\s_{b}\s_{a}^{-1})}{\prod_{a,b}^{N}(1-x\s_{a}\s_{b}^{-1})}
\frac{\prod_{a,b}^{N}\prod_{m=1}^{\infty}(1-\s_{a}\s_{b}^{-1}q^{m})(1-\s_{b}\s_{a}^{-1}q^{m})}{\prod_{a,b}^{N}\prod_{m=1}^{\infty}(1-x\s_{a}\s_{b}^{-1}q^{m})(1-x^{-1}\s_{b}\s_{a}^{-1}q^{m})},
\ea
and obeys by the relation \cite{Gaiotto:2021xce} 
\be
\la{3.4}
\sum_{N=0}^{\infty}\frac{\theta_{0}(ux^{N};q)}{\theta_{0}(u;q)}\I_{N}(x; q)x^{\frac{1}{2}N(N+1)}z^{N} = \prod_{m=-\infty}^{\infty}\bigg(1+\frac{zx^{m}}{1-u q^{m}}\bigg),
\ee
where $\theta_{0}(z; p) = \prod_{j=0}^{\infty}(1-zp^{j})(1-z^{-1}p^{j+1})$.
The index can be expanded in powers of $q$ 
\be
\I_{N}(x; q) = \sum_{m=0}^{\infty}\I_{N}^{(m)}(x)\,q^{m},
\ee
and the exact coefficient function $\I_{N}^{(m)}(x)$ can be determined by expanding (\ref{3.3}) in powers of $q$, and 
doing the integration over $|\s_{a}|=1$, assuming for instance $|x|<1$ to select the relevant poles. 
A more efficient algorithm is based on (\ref{3.4}). The key remark is that its r.h.s. can be expanded in powers of $q$ in exact form. 
For instance, at order $q^{0}$ and $q^{1}$ we have \footnote{Here, standard notation is used for the q-Pochhammer symbol 
 $(z; q)_{\infty}= \prod_{k=0}^{\infty}(1-z q^{k})$.}
\ba
\prod_{m=-\infty}^{\infty}\bigg(1+\frac{zx^{m}}{1-u q^{m}}\bigg) \bigg|_{q^{0}} &= \frac{1-u+z}{(1-u)(1+z)}(-z; x)_{\infty}, \\
\prod_{m=-\infty}^{\infty}\bigg(1+\frac{zx^{m}}{1-u q^{m}}\bigg) \bigg|_{q^{1}} &= -\frac{z(1-u+z)(1+xz-u^{2}x^{2})}{ux(1-u)(1+z)(1+xz)}(-z; x)_{\infty},
\ea
with similar expressions for higher powers of $q$. Comparing with the l.h.s. of (\ref{3.4}) gives $\I_{N}(x,q)$ at a certain order in $q$.
The terms with $N=1,2,3$ and $m=0,\dots, 5$
are explicitly 
\ba
\la{3.8}
\I_{1}(x; q) =& \frac{1}{1-x}+(x^{-1}-1)\,q+(x^{-2}-x)\,q^{2}+(x^{-3}-x^{-1}+1-x^{2})\,q^{3}+(x^{-4}-x^{3})\,q^{4}\lp
+(x^{-5}-x^{-2}+x-x^{4})\,q^{5}+\cdots,\\
\la{3.9}
\I_{2}(x; q) =& \frac{1}{(1-x)(1-x^{2})}+x^{-1} q+(2x^{-2}+x) 
q^2+(2x^{-3}+x^{-1}+x^3) q^3 \lp
+(3+3x^{-4}+x^5) q^4+(3x^{-5}+2 x+x^7) q^5+\cdots, \\
\la{3.10}
\I_{3}(x; q) =& \frac{1}{(1-x)(1-x^{2})(1-x^{3})}+\frac{1}{x 
(1-x^2)}\,q+\frac{(2+x-x^2+x^4)}{x^2 (1-x^2)}\,q^{2}\lp
+\frac{3-2x^{2}+2x^{3}+x^{4}-x^{6}+x^{8}}{x^{3}(1-x^{2})}\,q^{3}
+\frac{4+x-2 x^2+x^3+2 
x^6+x^7-x^{10}+x^{12}}{x^4 (1-x^2)}\,q^{4}\lp
+\frac{5-4 x^2+2 
x^3+x^4+x^6-2 x^7+2 x^9+x^{10}-x^{14}+x^{16}}{x^5 (1-x^2)}\,q^{5}+\cdots,
\ea
and so on. It is easy to obtain these expansion in $q$, but with the dependence on $x$ in closed form, for any higher $N$.
Notice that the $q^{0}$ and $q^{1}$ terms have the simple expressions
\be
\la{3.11}
\I_{N}^{(0)}(x) = \frac{1}{\prod_{m=1}^{N}(1-x^{m})}, \qquad \I_{N}^{(1)}(x) = -\frac{1-x}{x\,\prod_{m=1}^{N-1}(1-x^{m})}.
\ee
The brane expansion is (\ref{1.4})
where
\be
\la{3.12}
\I_{\infty}(x; q) = \prod_{m=1}^{\infty}\frac{1-q^{m}}{(1-x^{m})(1-x^{-m}q^{m})}.
\ee
Gaiotto-Lee suggested the analytic continuation formula for the brane indices
\be
\la{3.13}
\hat\I_{N}(x; q) = \I_{N}(x^{-1};  x^{-1}q).
\ee
Hence, we should have the following non-trivial relations for the finite $N$ indices
\ba
\la{3.14}
\I_{N}(x; q) =&   
\I_{\infty}(x; q)\sum_{k=0}^{\infty}x^{kN}\I_{k}(x^{-1}; x^{-1}q).
\ea
If we fix $N$ and increase the max value of $k$ in the sum, we can check that we have an equality. This is fully explicit for the coefficient of $q^{0}$ and $q^{1}$.
Indeed, using (\ref{3.11}) and 
\be
\I_{\infty}(x; q) = \frac{1}{(x;x)_{\infty}}-\frac{1-x}{x\,(x;x)_{\infty}}\,q+\mc O(q^{2}),
\ee  
the first two terms in the expansion of (\ref{3.14}) read
\ba
& \frac{1}{\prod_{m=1}^{N}(1-x^{m})} -\frac{1-x}{x\,\prod_{m=1}^{N-1}(1-x^{m})}\,q+\cdots =\sum_{k=0}^{\infty}x^{kN}\lp \bigg[\frac{1}{(x;x)_{\infty}}
-\frac{1-x}{x\,(x;x)_{\infty}}\,q+\cdots\bigg]
 \bigg[\frac{1}{\prod_{m=1}^{k}(1-x^{-m})} -\frac{1-x}{x\,\prod_{m=1}^{k-1}(1-x^{-m})}\,q+ \bigg]
\ea
One can check that this is satisfied if the following single condition holds
\be
\la{3.17}
\prod_{m=N+1}^{\infty}(1-x^{m}) = \sum_{k=0}^{\infty}x^{kN}\prod_{s=1}^{k}\frac{1}{1-x^{-s}} ,
\ee
which is easily proved. \footnote{We write the r.h.s. as 
$ \sum_{k=0}^{\infty}x^{kN}\prod_{s=1}^{k}\frac{-x^{s}}{1-x^{s}}  = 
 \sum_{k=0}^{\infty}(-1)^{k}x^{kN+k(k-1)/2}\prod_{s=1}^{k}\frac{1}{1-x^{s}}$.
Then, Euler identity
$\prod_{m=0}^{\infty}(1+q^{m}z) = \sum_{k=0}^{\infty}z^{k}\,q^{k(k-1)/2}\,\prod_{s=1}^{k}\frac{1}{1-q^{s}}$,
allows to transform it into the infinite product $\prod_{m=0}^{\infty}(1-x^{N+1}x^{m})$
which is same as the l.h.s. in (\ref{3.17}).}
Similarly, we can analyze  higher powers of $q$. 
To discuss the convergence of the brane expansion, we introduce the remainder 
 \be
 \Delta_{N}^{(K)}(x; q) = \I_{N}(x; q) -\I_{\infty}(x; q)\sum_{k=0}^{K}x^{kN}\I_{k}(x^{-1}; x^{-1}q).
 \ee
 Picking the leading non vanishing power of $x\to 0$, one finds \footnote{The exponent is for the first deviation term. It is thus equal to the
exponent in Eq.~(C.3) of \cite{Gaiotto:2021xce} increased by one.}
\be
\Delta_{N}^{(K)}(x; q) = \sum_{p=0}^{\infty}(c_{N,p}^{(K)}\, x^{\frac{1}{2}(K+1)(K+2+2N)-(K+2)\,p}+\cdots)\, q^{p},
\ee
that  clarifies to what extent (\ref{3.14}) holds. At fixed order in $q$, the agreement improves (the exponent of $x$ gets bigger) by adding more terms, \ie increasing $K$. On the other hand, with a fixed number
of terms, the leading correction has a leading contribution with a large negative exponent of $x$ as the order in $q$ is increased.

\subsection{Symmetric formulation}

As we mentioned in the Introduction, we want to work out a symmetric formulation in the fugacities $x,y$
and capture in particular the index and its brane expansion when
\be
x,y\to 0, \ \ \frac{x}{y} = \text{fixed}.
\ee
This is achieved by scaling 
\be
\la{3.21}
x\to \eps x, \qquad y\to \eps y,
\ee
where $\eps$ is a counting parameter. We will see that this is a non-trivial change since at a finite order in $\eps$ an infinite number of $q^{n}$ terms in 
the unsymmetric brane index $\hat\I_{k}(x;q)$ will contribute.

Let us begin by examining  the finite $N$ index in the $\eps$-expansion. Setting $q=xy$,
 from (\ref{3.4}) we read the expansion of $\I_{N}$ in symmetric homogenous polynomials
\ba
\la{3.22}
\I_{1}(\eps x,\eps y) &= 1+(x+y) \eps +(x^2-x y+y^2) \eps ^2+(x+y) (x^2-x y+y^2) 
\eps ^3+(x^4+y^4) \eps ^4+\cdots, \\
\la{3.23}
\I_{2}(\eps x, \eps y) &= 1+(x+y) \eps +2 (x^2+y^2) \eps ^2+2 (x+y) (x^2-x y+y^2) 
\eps ^3+3 (x^4+y^4) \eps ^4+\cdots, \\
\la{3.24}
\I_{3}(\eps  x, \eps y) &= 1+(x+y) \eps +2 (x^2+y^2) \eps ^2+(x+y) (3 x^2-2 x y+3 y^2) 
\eps ^3+(4 x^4+x^2 y^2+4 y^4) \eps ^4+\cdots.
\ea
Also, we have 
\ba
\I_{\infty}(\eps x, \eps y) &=1+(x+y) \eps +2 (x^2+y^2) \eps ^2+(x+y) (3 x^2-2 x y+3 y^2) 
\eps ^3\lp
+(5 x^4+x^3 y+2 x^2 y^2+x y^3+5 y^4) \eps ^4+\cdots.
\ea
These expressions are clearly equivalent to the previous expressions for $\I_{N}(x,q)$ in (\ref{3.8}), (\ref{3.9}), and (\ref{3.10}). For instance, in (\ref{3.8})
we can replace $q=xy$ and get 
\ba
\la{3.26}
\I_{1}(x; xy) =& \frac{1}{1-x}+(x^{-1}-1)\,xy+(x^{-2}-x)\,x^{2}y^{2}+(x^{-3}-x^{-1}+1-x^{2})\,x^{3}y^{3}\lp
+(x^{-4}-x^{3})\,x^{4}y^{3}+(x^{-5}-x^{-2}+x-x^{4})\,x^{5}y^{5}+\mc O(y^{6}),
\ea
and the written terms agree with all terms in (\ref{3.22}) neglecting powers of $y$ higher then 5. In particular, the term $y^{0}$ in (\ref{3.26}) resum all $x^{n}$ terms in (\ref{3.22}).
For higher $N$ similar resummations occur and in general using the exact depencence on $x$ in $\I_{N}(x,xy)$ resums terms of the form $x^{n}y^{p}$ with fixed $p$ and any $n$. 

\medskip
The natural brane expansion for the index $\I_{N}(x,y)$ is expected to be 
\ba
\la{3.27}
\I_{N}(x,y) &= \I_{\infty}(x,y)\,\bigg[1+\sum_{k,k'=1}^{\infty}x^{kN}y^{k'N}\hat\I_{(k,k')}(x,y)\bigg],
\ea
where \cf (\ref{3.12}), 
\be
\I_{\infty}(x,y) = \frac{(xy; xy)_{\infty}}{(x;x)_{\infty}(y;y)_{\infty}}.
\ee
In the next section, we will compute $\hat I_{(k,k')}$ by multivariate residue computations.

\subsection{Multivariate residue computation of the brane expansion}

Setting $q=xy$ in (\ref{3.2}), the  single-letter index becomes symmetric in $x,y$
\be
f(x,y) = \frac{x+y-2xy}{1-xy},
\ee
and using  (\ref{2.5}) we obtain
\ba
\hat f^{X}_{X} &= \frac{x^{-1}-2y+xy}{1-y} = \sum_{m=0}^{\infty}(x^{-1}-2y+xy)y^{m}, \\
\hat f^{Y}_{Y} &= \frac{y^{-1}-2x+xy}{1-x} = \sum_{m=0}^{\infty}(y^{-1}-2x+xy)x^{m}, \\
\hat f^{X}_{Y} &=y^{-1}-x, \qquad
\hat f^{Y}_{X} =x^{-1}-y. 
\ea
We now want to compute the terms in the r.h.s. of (\ref{3.27}).  From (\ref{2.4}), the general formula for $\hat\I_{(k,k')}$ is 
\ba
\hat\I_{(k,k')}(x,y) &= \frac{1}{k!k'!}\oint
\prod_{a=1}^{k}\frac{d\s_{a}^{X}}{2\pi i \s_{a}^{X}}
\prod_{a=1}^{k'}\frac{d\s_{a}^{Y}}{2\pi i \s_{a}^{Y}}
\prod_{a\neq b}^{k}(1-\s^{X}_{a}/\s^{X}_{b})
\prod_{a\neq b}^{k'}(1-\s^{Y}_{a}/\s^{Y}_{b})\lp
\prod_{a,b=1}^{k}\prod_{m=0}^{\infty}\frac{(1-y^{m+1}\s_{a}^{X}/\s_{b}^{X})^{2}}{(1-x^{-1}y^{m}\s_{a}^{X}/\s_{b}^{X})
(1-x y^{m+1}\s_{a}^{X}/\s_{b}^{X})}\lp
\prod_{a,b=1}^{k'}\prod_{m=0}^{\infty}\frac{(1-x^{m+1}\s_{a}^{Y}/\s_{b}^{Y})^{2}}{(1-y^{-1}x^{m}\s_{a}^{Y}/\s_{b}^{Y})
(1-y x^{m+1}\s_{a}^{Y}/\s_{b}^{Y})}\lp
\prod_{a=1}^{k}\prod_{b=1}^{k'}\frac{1-x\s_{a}^{X}/\s_{b}^{Y}}{1-y^{-1}\s_{a}^{X}/\s_{b}^{Y}}
\frac{1-y\s_{b}^{Y}/\s_{a}^{X}}{1-x^{-1}\s_{b}^{Y}/\s_{a}^{X}}.
\ea
The exchange symmetry
\be
\hat\I_{(k,k')}(x,y) = \hat\I_{(k',k)}(y,x),
\ee
allows to consider  just the cases $k\ge k'$.
The sum $k+k'$ will be called the level of the brane index.

\subsubsection{Level 1}

At this level we need just to compute the $(1,0)$ contribution. It does not require any integration and has the exact expression 
\ba
\hat\I_{(1,0)}(x,y) &=
\prod_{m=0}^{\infty}\frac{(1-y^{m+1})^{2}}{(1-x^{-1}y^{m})
(1-x y^{m+1})} = \frac{(y;y)^{2}_{\infty}}{(x^{-1};y)_{\infty}(xy;y)_{\infty}}.
\ea
Its small $\eps$ expansion after the scaling (\ref{3.21})  is
\ba
\hat\I_{(1,0)}&(\eps x, \eps y) =
 -\frac{x^2 \eps }{x-y}-x (x-y) \eps ^2+(-x^3+y^3) \eps^3\lp
+\bigg(-x^4+x^2 y^2-x y^3+\frac{y^5}{x}\bigg) \eps ^4+\bigg(-x^{5}+\frac{y^{7}}{x^{2}}\bigg)\eps^{5}
+\bigg(-x^6+x^3 y^3-y^6+\frac{y^9}{x^3}\bigg) \eps ^6+\cdots.
\ea
The $\mc O(\eps)$ term is highly non-trivial since it may be expanded in small $x/y$ or $y/x$ with different results. The singularity at the codimension-1 wall $x=y$
has to cancel in the full index which has no such wall-crossing problems.

\subsubsection{Level 2}

At level 2 we need  $(2,0)$ and $(1,1)$.
The expression for $(2,0)$ is 
\ba
\hat\I_{(2,0)}(x,y) &= \frac{1}{2}\oint
\prod_{a=1}^{2}\frac{d\s_{a}}{2\pi i \s_{a}}
\prod_{a\neq b}^{2}(1-\s_{a}/\s_{b})
\prod_{a,b=1}^{2}\prod_{m=0}^{\infty}\frac{(\s_{b}-y^{m+1}\s_{a})^{2}}{(\s_{b}-x^{-1}y^{m}\s_{a})
(\s_{b}-x y^{m+1}\s_{a})} \lp
= \frac{1}{2}\prod_{m=0}^{\infty}\frac{(1-y^{m+1})^{4}}{(1-x^{-1}y^{m})^{2}(1-x y^{m+1})^{2}}\oint
\prod_{a=1}^{2}\frac{d\s_{a}}{2\pi i \s_{a}}
\prod_{a\neq b}^{2}(1-\s_{a}/\s_{b})
\prod_{m=0}^{\infty}\frac{(\s_{b}-y^{m+1}\s_{a})^{2}}{(\s_{b}-x^{-1}y^{m}\s_{a})
(\s_{b}-x y^{m+1}\s_{a})}
\ea
This can be computed by setting $\s_{2}=1$ and summing over the poles in $\s_{1}$ in $g_{1}$, see Appendix B of \cite{Lee:2022vig} The result is 
\ba
\la{3.38}
\hat\I_{(2,0)} &= -\frac{x^7 (x-2 y) \eps ^4}{(x-y)^2 y (x+y)}+\bigg(2 x^5-\frac{x^7}{y^2}\bigg) 
\eps ^5+\bigg(2 x^6-\frac{x^9}{y^3}+2 x^3 y^3\bigg) \eps ^6+\bigg(3 
x^7-\frac{x^{11}}{y^4}+2 x y^6\bigg) \eps ^7\lp
+\bigg(3 x^8-\frac{x^{13}}{y^5}+3 
x^4 y^4+\frac{2 y^9}{x}\bigg) \eps ^8+\bigg(4 x^9-\frac{x^{15}}{y^6}+x^3 
y^6+\frac{2 y^{12}}{x^3}\bigg) \eps ^9+\cdots.
\ea
Then, let us consider $(1,1)$
\ba
\hat\I_{(1,1)}(x,y) &=
\prod_{m=0}^{\infty}\frac{(1-y^{m+1})^{2}}{(1-x^{-1}y^{m})(1-x y^{m+1})}
\frac{(1-x^{m+1})^{2}}{(1-y^{-1}x^{m})(1-y x^{m+1})}\lp
\oint
\frac{d\s^{X}}{2\pi i \s^{X}}
\frac{d\s^{Y}}{2\pi i \s^{Y}}
\frac{\s^{Y}-x\s^{X}}{\s^{Y}-y^{-1}\s^{X}}
\frac{\s^{X}-y\s^{Y}}{\s^{X}-x^{-1}\s^{Y}}.
\ea
The multivariate residue is computed again setting $\s^{Y}=1$ and summing over the poles in $\s^{X}$.  The result is  
\ba
\hat\I_{(1,1)}(x,y) =&
xy \prod_{m=0}^{\infty}\frac{(1-y^{m+1})^{2}}{(1-x^{-1}y^{m})(1-x y^{m+1})}
\frac{(1-x^{m+1})^{2}}{(1-y^{-1}x^{m})(1-y x^{m+1})} \lp
= \frac{xy\,(x;x)^{2}_{\infty}\,(y;y)^{2}_{\infty}}{(x^{-1};y)_{\infty}\,(y^{-1};x)_{\infty}\,(xy;x)_{\infty}\,(xy;y)_{\infty}}.
\ea
After the scaling (\ref{3.21})  we obtain 
\ba
\hat\I_{(1,1)}(\eps x, \eps y) =&
 -\frac{x^3 y^3 \eps ^4}{(x-y)^2}-x^2 y^2 (x+y) \eps ^5-x y (x^4+2 x^3 
y+2 x y^3+y^4) \eps ^6\lp
+(-x^7-2 x^6 y-x^4 y^3-x^3 y^4-2 x y^6-y^7) 
\eps ^7+\cdots.
\ea

\subsubsection{Level 3}

At level 3 we need  $(3,0)$ and $(2,1)$. The first is 
\ba
\hat\I_{(3,0)}(x,y) &= \frac{1}{3!}\oint
\prod_{a=1}^{3}\frac{d\s_{a}}{2\pi i \s_{a}}
\prod_{a\neq b}^{3}(1-\s_{a}/\s_{b})
\prod_{a,b=1}^{3}\prod_{m=0}^{\infty}\frac{(\s_{b}-y^{m+1}\s_{a})^{2}}{(\s_{b}-x^{-1}y^{m}\s_{a})
(\s_{b}-x y^{m+1}\s_{a})} \lp
= \frac{1}{3!}\prod_{m=0}^{\infty}\frac{(1-y^{m+1})^{6}}{(1-x^{-1}y^{m})^{3}(1-x y^{m+1})^{3}}\oint
\prod_{a=1}^{3}\frac{d\s_{a}}{2\pi i \s_{a}}
\prod_{a\neq b}^{3}(1-\s_{a}/\s_{b})
\prod_{m=0}^{\infty}\frac{(\s_{b}-y^{m+1}\s_{a})^{2}}{(\s_{b}-x^{-1}y^{m}\s_{a})
(\s_{b}-x y^{m+1}\s_{a})}
\ea
A long calculation with the deformation algorithm gives the first term in the small $\eps$ expansion after the scaling (\ref{3.21}) 
\ba
\hat\I_{(3,0)}(\eps x, \eps y) &= -\frac{x^{15}\,(2x^{3}-3x^{2}y-3xy^{2}+5y^{3})}{y^{3}(x-y)^{3}(x+y)(x^{2}+xy+y^{2})}\,\eps^{9}+\mc O(\eps^{10}).
\ea
In the $(2,1)$ case we have 
\ba
\hat\I_{(2,1)}(x,y) &= \frac{1}{2!}\prod_{m=0}^{\infty}\frac{(1-x^{m+1})^{2}(1-y^{m+1})^{4}}{(1-y^{-1}x^{m})(1-y x^{m+1})(1-x^{-1}y^{m})^{2}(1-x y^{m+1})^{2}}\lp
\oint
\prod_{a=1}^{2}\frac{d\s_{a}^{X}}{2\pi i \s_{a}^{X}}
\frac{d\s^{Y}}{2\pi i \s^{Y}}
\prod_{a\neq b}^{2}(1-\s^{X}_{a}/\s^{X}_{b})
\prod_{a=1}^{2}\frac{\s^{Y}-x\s_{a}^{X}}{\s^{Y}-y^{-1}\s_{a}^{X}}
\frac{\s_{a}^{X}-y\s^{Y}}{\s_{a}^{X}-x^{-1}\s^{Y}}\lp
\prod_{a\neq b}^{2}\prod_{m=0}^{\infty}\frac{(\s_{b}^{X}-y^{m+1}\s_{a}^{X})^{2}}{(\s_{b}^{X}-x^{-1}y^{m}\s_{a}^{X})
(\s_{b}^{X}-x y^{m+1}\s_{a}^{X})}.
\ea
We notice that we can integrate out $\s^{Y}$ by a rescaling. Following Lee's prescription in the rest (where we simply set $\s^{Y}=1$) we define
\ba
g_{1} &= (\s_{1}^{X})^{2}(\s_{1}^{X}-x^{-1})(\s_{1}^{X}-y)\prod_{m=0}^{\infty}(\s_{1}^{X}-x^{-1}y^{m}\s_{2}^{X})
(\s_{1}^{X}-x y^{m+1}\s_{2}^{X}), \\
g_{2} &= (\s_{2}^{X})^{2}(\s_{2}^{X}-x^{-1})(\s_{2}^{X}-y)\prod_{m=0}^{\infty}(\s_{2}^{X}-x^{-1}y^{m}\s_{1}^{X})
(\s_{2}^{X}-x y^{m+1}\s_{1}^{X}).
\ea
Using again the deformation algorithm, we compute the first term of the small $\eps$ expansion after the scaling (\ref{3.21})
\be
\hat\I_{(2,1)}(\eps x, \eps y) = -\frac{x^{9}y^{3}(x-2y)}{(x-y)^{3}(x+y)}\eps^{9}+\cdots.
\ee

\subsection{Cancellation of wall-crossing poles}

The first term in the computed brane indices $\hat\I_{(k,k')}$ is a non-trivial rational function of the fugacities
with wall-crossing pole at $x=\pm y$. Let us show that these poles cancel in the sum of contributions at given level
for any $N$. This has to happen since the index is a polynomial in $x,y$.

\paragraph{level 1}

The non-trivial rational functions at level 1 contribute in (\ref{3.27}) as 
\be
\la{3.48}
-x^{N}\frac{x^{2}}{x-y}-y^{N}\frac{y^{2}}{y-x} = -\frac{x^{N+2}-y^{N+2}}{x-y} = -x^{N+1}\frac{1-(y/x)^{N+2}}{1-y/x}.
\ee
and this is a polynomial.

\paragraph{level 2}
At level 2, we need
\ba
& -x^{2N}\frac{x^{7}(x-2y)}{(x-y)^{2}y(x+y)}-(xy)^{N}\frac{x^{3}y^{3}}{(x-y)^{2}}-y^{2N}\frac{y^{7}(y-2x)}{(x-y)^{2}x(x+y)}\lp
= -\frac{(x^{N+4}-y^{N+4})(x^{N+5}-2x^{N+4}y+2xy^{N+4}-y^{N+5})}{xy(x-y)^{2}(x+y)}
\ea
and this is $1/(xy)$ times a polynomial, because one checks that the limits $x\to \pm y$ are not singular. So we get only simple monomials and no wall-crossing denominators.

\paragraph{level 3}
We have to consider the combination 
\be
x^{3N}\hat\I_{(3,0)}(x,y)+
x^{2N}y^{N}\hat\I_{(2,1)}(x,y)+
x^{N}y^{2N}\hat\I_{(2,1)}(y,x)+
y^{3N}\hat\I_{(3,0)}(y,x).
\ee
Using the previous expressions we can check that for all $N$ this is a finite sum of monomials in $x^{\pm 1},y^{\pm 1}$, hence again all wall-crossing unwanted
denominators cancel. 

\subsection{Checking the validity of the symmetric brane expansion}

Let us see how the brane expansion reproduces the exact index, \ie the validity of the relation (\ref{3.27}). Introducing the explicit scaling (\ref{3.21}), it reads
\ba
\I_{N}(\eps x, \eps y) &= \I_{\infty}(\eps x, \eps y)\,\bigg[1+\sum_{k,k'=1}^{\infty}\eps^{(k+k')N}\,
x^{kN}y^{k'N}\hat\I_{(k,k')}(\eps x,\eps y)\bigg].
\ea
Let us remark that up to level 2
we have exact expression for all terms with the exception of $\hat\I_{(2,0)}$ that can be expanded in $\eps$ with minor effort. At level 3, we have 
the leading $\eps^{9}$ result for $\hat\I_{(3,0)}(\eps x, \eps y)$ and $\hat\I_{(2,1)}(\eps x, \eps y)$. This means that for a generic $N$, we can appreciate the role
of the computed terms, in particular the level 3 contributions, by computing the exact index at order $\eps^{3N+9}$ and $\hat\I_{(2,0)}$ at order 
$\eps^{N+9}$. The necessary expansions for making checks at $N=1$ and $N=2$, extending the partial results quoted in (\ref{3.22}) , (\ref{3.23}) , and (\ref{3.38}) are
given below. While the expansion of $\I_{1}$ and $\I_{2}$ is somewhat trivial, the one for $\hat\I_{(2,0)}$ is not. Reason is that to compute correctly the higher terms in the $\eps$-expansion, 
we need to increase the number of factors kept in the infinite products in the integrand expression. Our results are
\ba
\I_{1}& (\eps x, \eps y) = 1+(x+y) \eps +(x^2-x y+y^2) \eps ^2+(x+y) (x^2-x y+y^2) \eps 
^3+(x^4+y^4) \eps ^4\lp
+(x-y)^2 (x+y) (x^2+x y+y^2) \eps ^5+(x^6+x^3 
y^3+y^6) \eps ^6
+(x+y) (x^6-x^5 y+x^4 y^2\lp
-x^3 y^3+x^2 y^4-x y^5+y^6) 
\eps ^7+(x-y)^2 (x^2+x y+y^2)
 (x^4+x^3 y+x^2 y^2+x y^3+y^4) \eps 
^8\lp
+(x+y) (x^2-x y+y^2) (x^6-x^3 y^3+y^6) \eps ^9
+(x^2+y^2) (x^8-x^6 
y^2+x^4 y^4-x^2 y^6+y^8) \eps ^{10}\lp
+(x+y) (x^{10}-x^9 y+x^8 y^2-x^7 
y^3
+x^5 y^5-x^3 y^7+x^2 y^8-x y^9+y^{10}) \eps ^{11}+(x^{12}-x^6 
y^6+y^{12}) \eps ^{12}+\cdots, 
\ea
\ba
\I_{2}& (\eps x, \eps y) =
1+(x+y) \eps +2 (x^2+y^2) \eps ^2+2 (x+y) (x^2-x y+y^2) \eps ^3+3 
(x^4+y^4) \eps ^4\lp
+(x+y) (3 x^4-3 x^3 y+4 x^2 y^2-3 x y^3+3 y^4) \eps 
^5+4 (x^2+y^2) (x^4-x^2 y^2+y^4) \eps ^6\lp
+4 (x+y) (x^6-x^5 y+x^4 
y^2-x^3 y^3+x^2 y^4-x y^5+y^6) \eps ^7+(5 x^8+3 x^4 y^4+5 y^8) \eps 
^8\lp
+(x+y) (x^2-x y+y^2) (5 x^6-4 x^3 y^3+5 y^6) \eps ^9+6 (x^2+y^2) 
(x^8-x^6 y^2+x^4 y^4-x^2 y^6+y^8) \eps ^{10}\lp
+2 (x+y) (3 x^{10}-3 x^9 
y+3 x^8 y^2-3 x^7 y^3+3 x^6 y^4-2 x^5 y^5+3 x^4 y^6-3 x^3 y^7+3 x^2 
y^8-3 x y^9\lp
+3 y^{10}) \eps ^{11}+7 (x^4+y^4) (x^8-x^4 y^4+y^8) \eps 
^{12}+(x+y) (7 x^{12}-7 x^{11} y+7 x^{10} y^2-7 x^9 y^3 \lp
+8 x^8 y^4-8 
x^7 y^5+8 x^6 y^6-8 x^5 y^7+8 x^4 y^8-7 x^3 y^9+7 x^2 y^{10}-7 x 
y^{11}+7 y^{12}) \eps ^{13}\lp
+4 (x^2+y^2) (2 x^{12}-2 x^{10} y^2+2 x^8 
y^4-x^6 y^6+2 x^4 y^8-2 x^2 y^{10}+2 y^{12}) \eps ^{14}+8 (x+y) \lp
(x^2-x y+y^2) (x^4-x^3 y+x^2 y^2-x y^3+y^4) (x^8+x^7 y-x^5 y^3-x^4 
y^4-x^3 y^5+x y^7+y^8) \eps ^{15}+\cdots, 
\ea
and finally 
\ba
\hat\I_{(2,0)} & (\eps x, \eps y) =
-\frac{x^7 (x-2 y) \eps ^4}{(x-y)^2 y (x+y)}+\bigg(2 x^5-\frac{x^7}{y^2}\bigg) 
\eps ^5+\bigg(2 x^6-\frac{x^9}{y^3}+2 x^3 y^3\bigg) \eps ^6\lp
+\bigg(3 
x^7-\frac{x^{11}}{y^4}+2 x y^6\bigg) \eps ^7+\bigg(3 x^8-\frac{x^{13}}{y^5}+3 
x^4 y^4+\frac{2 y^9}{x}\bigg) \eps ^8 \lp
+\bigg(4 x^9-\frac{x^{15}}{y^6}+x^3 
y^6+\frac{2 y^{12}}{x^3}\bigg) \eps ^9+\bigg(4 x^{10}-\frac{x^{17}}{y^7}+3 x^5 
y^5+4 x^2 y^8+\frac{2 y^{15}}{x^5}\bigg) \eps ^{10}\lp
+\bigg(5 
x^{11}-\frac{x^{19}}{y^8}+2 x y^{10}+\frac{2 y^{18}}{x^7}\bigg) \eps 
^{11}+\cdots.
\ea
Let us now introduce the difference between the exact index and the approximate brane expansion 
corresponding to keeping terms up to level $\ell$ 
\be
\Delta_{N}^{(\ell)} = -\I_{N}(\eps x, \eps y) + \I_{\infty}(\eps x, \eps y)\,\bigg[1+\mathop{\sum_{k,k'=1}^{\infty}}_{k+k'\le \ell}\eps^{(k+k')N}\,
x^{kN}y^{k'N}\hat\I_{(k,k')}(\eps x,\eps y)\bigg].
\ee
At $N=1$, the level 1 difference is 
\be
\Delta_{N=1}^{(1)} = \bigg(-x^6+\frac{x^7}{y}-2 x^4 y^2-x^3 y^3-2 x^2 y^4-y^6+\frac{y^7}{x}\bigg)\,\eps^{6}+\cdots.
\ee
Including the level 2 terms it improves to 
\ba
\Delta_{N=1}^{(2)} =& \bigg(2 x^{12}+\frac{2 x^{15}}{y^3}-\frac{x^{14}}{y^2}-\frac{2 x^{13}}{y}-2 
x^{11} y-x^{10} y^2+2 x^9 y^3+x^8 y^4+5 x^6 y^6+x^4 y^8+2 x^3 y^9-x^2 
y^{10} \lp
-2 x y^{11}+2 y^{12}-\frac{2 
y^{13}}{x}-\frac{y^{14}}{x^2}+\frac{2 y^{15}}{x^3}\bigg)\,\eps^{12}+\cdots.
\ea
This is canceled by the level 3 terms leaving
\ba
\Delta_{N=1}^{(3)} =& \mc O(\eps^{20}),
\ea
where $20=16+4\times (N=1)$, coming from the expected leading powers of $\eps$ in $\hat\I$ at level 4 plus the $N$ dependence of that contribution in the brane expansion.

Similarly, at $N=2$ we find
\ba
\Delta_{N=2}^{(1)} = \bigg(-x^8+\frac{x^9}{y}-2 x^6 y^2-x^5 y^3-3 x^4 y^4-x^3 y^5-2 x^2 
y^6-y^8+\frac{y^9}{x}\bigg)\,\eps^{8}+\cdots.
\ea
At level 2, this is reduced to 
\ba
\Delta_{N=2}^{(2)} = & \bigg(
-2 x^{15}-\frac{2 x^{18}}{y^3}+\frac{x^{17}}{y^2}+\frac{2 
x^{16}}{y}+2 x^{14} y+x^{13} y^2-2 x^{12} y^3-x^{10} y^5-5 x^9 y^6 \lp
-2 
x^8 y^7-2 x^7 y^8-5 x^6 y^9-x^5 y^{10}-2 x^3 y^{12}+x^2 y^{13}+2 x 
y^{14}-2 y^{15}+\frac{2 y^{16}}{x}+\frac{y^{17}}{x^2}-\frac{2 
y^{18}}{x^3}
\bigg)\,\eps^{15}+\cdots,
\ea
and this is canceled by the level 3 terms leading to 
\ba
\Delta_{N=2}^{(3)} =& \mc O(\eps^{24}),
\ea
where $24=16+4\times (N=2)$. The above calculations show that the double brane expansion (\ref{3.27}) works indeed as expected
and reproduces the finite $N$ indices when both fugacities are sent to zero independently.

\section{Comparing the Gaiotto-Lee and symmetric expansions}
\la{sec:comp}

As we mentioned in the Introduction, we have two representations of the finite $N$ index and from (\ref{3.14}) and (\ref{3.27})
it should be that
\be
\sum_{k=1}^{\infty}x^{kN}\hat\I_{k}(x; q) = \sum_{k,k'=1}^{\infty}x^{kN}y^{k'N}\hat\I_{(k,k')}(x,y).
\ee
Using (\ref{3.13}), the l.h.s. can be written
\be
\sum_{k=1}^{\infty}x^{kN}\I_{k}(x^{-1}; x^{-1}q) = \sum_{k,k'=1}^{\infty}x^{kN}y^{k'N}\hat\I_{(k,k')}(x,y).
\ee
Finally, since $q=xy$, the relation we need to prove reads
\be
\la{4.3}
\sum_{k=1}^{\infty}x^{kN}\I_{k}(x^{-1}; y) = \sum_{k,k'=1}^{\infty}x^{kN}y^{k'N}\hat\I_{(k,k')}(x,y).
\ee
Here, we remark that the l.h.s. is known as a power series in $y$. Instead the $y$ dependence of the r.h.s. is non-trivial.
It is puzzling that (\ref{4.3}) could hold. To understand what is going on, let us inspect the $k=1$ term in the 
l.h.s. of (\ref{4.3}). From  (\ref{3.8}), it is 
\ba
& \eps^{N}x^{N}\I_{1}(\eps^{-1}x^{-1}; \eps y) \lp
= \eps^{N}x^{N}\bigg[ \frac{1}{1-\eps^{-1} x^{-1}}+(\eps x-1)\,\eps y+(\eps^{2}x^{2}-\eps^{-1}x^{-1})\,\eps^{2}y^{2}
+(\eps^{3}x^{3}-\eps x+1-\eps^{-2}x^{-2})\,\eps^{3}y^{3}\lp
+(\eps^{4}x^{4}-\eps^{-3}x^{-3})\,\eps^{4}y^{4}
+(\eps^{5}x^{5}-\eps^{2}x^{2}+\eps^{-1}x^{-1}-\eps^{-4}x^{-4})\,\eps^{5}y^{5}+\cdots\bigg] \lp
=  \eps^{N+1}x^{N}\bigg(-x-y-\frac{y^{2}}{x}-\frac{y^{3}}{x^{2}}-\frac{y^{4}}{x^{3}}+\mc O(\eps)\cdots\bigg)
\ea
As we remarked, at this fixed order in $\eps$, we receive contributions from all the $q^{n}$ terms in the unsymmetric index $\I_{1}$. 
Since the structure suggests a simple geometric series, we sum it up and get 
\ba
 \eps^{N}x^{N}\I_{1}(\eps^{-1}x^{-1}; \eps y)  =  -\eps^{N+1}x^{N+1}\bigg(\frac{1}{1-\frac{y}{x}}+\mc O(\eps)\bigg).
\ea
Comparing with (\ref{3.48}), we see that this equals the $(1,0)$ contribution in the symmetric brane expansion. This suggests that we may have in general
\be
\la{4.6}
\I_{k}(x^{-1}; y) = \hat\I_{(k,0)}(x,y).
\ee
Let us test the conjectured relation (\ref{4.6}) at level 2. From (\ref{3.9}), we obtain 
\ba
\I_{2}(\eps^{-1}x^{-1}; \eps y) &= \cdots +\frac{y^{5}}{\eps^{2}x^{7}}+\frac{y^{4}}{\eps x^{5}}+\frac{y^{3}}{x^{3}}
+\frac{y^{2}}{x}\eps+xy\eps^{2}+x^{3}\eps^{3}\lp
+\bigg(x^{4}+2x^{2}y^{2}+xy^{3}+3y^{4}+2\frac{y^{5}}{x}+\cdots\bigg)\,\eps^{4}+\mc O(\eps^{5})
\ea
The terms that are singular for $\eps\to 0$ appear to form again a geometric series that we sum
\ba
\cdots +\frac{y^{5}}{\eps^{2}x^{7}}+\frac{y^{4}}{\eps x^{5}}+\frac{y^{3}}{x^{3}}
+\frac{y^{2}}{x}\eps = \eps\frac{y^{2}}{x}\frac{1}{1-\frac{y}{\eps x^{2}}}
\ea
and in this form we can re-expand at small $\eps$ to get 
\be
 \eps\frac{y^{2}}{x}\frac{1}{1-\frac{y}{\eps x^{2}}} = -\eps^{2}xy-\eps^{3}x^{3}-\eps^{4}\frac{x^{5}}{y}+\mc O(\eps^{5}).
 \ee
 Hence, the small $\eps$ expansion is 
\ba
\I_{2}(\eps^{-1}x^{-1}; \eps y) &= \bigg(-\frac{x^{5}}{y}+x^{4}+2x^{2}y^{2}+xy^{3}+3y^{4}+2\frac{y^{5}}{x}+\cdots\bigg)\,\eps^{4}+\mc O(\eps^{5}).
\ea
and this precisely agrees with the small $y$ expansion of the $\eps^{4}$ term in $\hat\I_{(2,0)}(x,y)$
\be
-\frac{x^{7}(x-2y)}{(x-y)^{2}y(x+y)} = -\frac{x^{5}}{y}+x^{4}+2x^{2}y^{2}+xy^{3}+3y^{4}+2\frac{y^{5}}{x}+4\frac{y^{6}}{x^{2}}+3\frac{y^{7}}{x^{3}}
+\mc O(y^{8}).
\ee
Thus, at level 2, the relation (\ref{4.6}) holds although its verification is non-trivial, since we had to resum the singular terms for $\eps\to 0$.
A similar analysis may be attempted at level 3, but as expected, the resummation of the singular terms is less trivial to be guessed.

The above analysis and in particular the non-trivial relation (\ref{4.6}) shows that one can restrict the symmetric brane expansion to the form 
\ba
\I_{N}(x,y) &= \I_{\infty}(x,y)\,\bigg[1+\sum_{k=1}^{\infty}x^{kN}\hat\I_{(k,0)}(x,y)\bigg],
\ea
provided we expand in an asymmetric way the r.h.s. by first expanding in $y$ and then in $x$. A similar mechanisms holds for a  simplified index  of the 3d SCFTs associated to M2 branes
as discussed in the Introduction of \cite{Gaiotto:2021xce}. To appreciate what is the fate of the missing terms $(k,k')$  with $k'>0$, it is instructive to look at  the expression multiplying $y^{k'N}$, \ie
\be
\hat\I_{(0,1)}(x,y) = \frac{(x;x)^{2}_{\infty}}{(y^{-1};x)_{\infty}(xy;x)_{\infty}}.
\ee
When we replace $y=q/x$, the factor $(x/q; x)^{-1}_{\infty}$ is zero at all orders in the expansion in series of  $q$ at fixed $x$. This suggests that
all terms $(k,k')$ with $k'>0$ are invisible if this order of expansions is applied, while they are fully relevant in the (symmetric) $\eps$-expansion.

\section{Enhanced degeneracies at the walls}
\la{sec:deg}

We conclude with some comments on a question posed in \cite{Lee:2022vig}, \ie 
whether we can find degeneracies in brane indices consistent with  excited geometries,  as BPS black holes or bubbling geometries that break more supersymmetry.
In that paper, it was claimed that such degeneracies 
emerge as a finite remainder left after a cancellation of poles happens at wall-crossing points where two or more fugacities collapse or 
satisfy some special relation. 
Such a process of cancellation among contributions coming from different brane configurations 
to produce degeneracies associated to objects such as black holes or bubbling geometries 
could be interpreted as a signal of a process by which the corresponding stacks of branes bind and form bound states.
For example, a representative example of one such cancellation happens in  the giant graviton expansion of the 
$\frac{1}{16}$-BPS index that in the coincident limit $x = y = z = w^{2}$ and $p=q=w^{3}$, \cf (\ref{3.1}), 
takes the form
\be
\la{5.1}
\I_{N} = \I_{\infty}\Big[1+\sum_{n=1}^{\infty}w^{2nN}\sum_{p=0}^{\infty}D_{n,p}(N)w^{2n^{2}+p}\Big],
\ee
where $D_{n,p}(N)$ is a polynomial of degree $3n-1$ in $N$  \cite{Lee:2022vig}.

In the present case of the Schur index, it is interesting to explore this phenomenon in the $\eps$-expansion
and compare with the known giant-graviton expansion at the special point $x=y$, \ie for the $\frac{1}{8}$-BPS index. This reads \cite{Bourdier:2015wda}
\be
\la{52}
\frac{\I_{N}(x, x)}{\I_{\infty}(x, x)} = \sum_{k=0}^{\infty}(-1)^{k}\bigg[\binom{N+k}{N}+\binom{N+k-1}{N}\bigg]\,x^{kN+k^{2}},
\ee
and is indeed characterized by $N$-dependent degeneracies. The terms in (\ref{52}) can be understood from our expressions as we know explain. At 
level 1 we have the following contribution to the ratio in the l.h.s. of (\ref{52})
\be
\eps^{N}\bigg[x^{N}\hat\I_{(1,0)}(\eps x, \eps y)+y^{N}\hat \I_{(0,1)}(\eps x, \eps y)\bigg].
\ee
 Expanding in $\eps$ and taking the non-singular limit $y\to x$ gives a single term.
 \be
 -(N+2)\eps^{N+1}x^{N+1}.
 \ee
 Doing the same calculation at level 2 we get again a single contribution
 \be
 \frac{1}{2}(N+4)(N+1)\eps^{2N+4}x^{2N+4}.
 \ee
 Finally, at level 3, we have computed just the first non vanishing contribution to the various $\hat\I_{(k,k')}$ with $k+k'=3$ and from those terms
 we obtain 
 \be
 -\frac{1}{6}(N+6)(N+2)(N+1)\eps^{3N+9}x^{3N+9}.
 \ee
 One can indeed check that these contributions match the $k=1,2,3$ terms in (\ref{52}).

\section*{Acknowledgements}
We thank J. H. Lee and Y. Imamura for useful discussions. We also acknowledge financial support 
from the INFN grant GSS (Gauge Theories, Strings and Supergravity).

\appendix

\section{$\eps$-expansion in the $\frac{1}{4}$-BPS sector}
\la{app:quart}

It is instructive to consider the $\eps$-expansion of the index for the $\frac{1}{4}$-BPS sector. The single-letter index depends on 
two fugacities since it is obtained from (\ref{3.1}) by sending $p,q,z\to 0$. It reads
\be
f = x+y-xy.
\ee
The full index has generating function \cite{Kinney:2005ej}
\be
\la{B.2}
\sum_{N=0}^{\infty}\zeta^{N}\I_{N}(x,y) = \frac{1}{1-\zeta}\prod_{n=1}^{\infty}\frac{1}{(1-\zeta x^{n})(1-\zeta y^{n})}.
\ee
Taking residues in $\zeta$ one obtains the brane expansion.
This gives
\ba
\I_{N} &= \I_{\infty}\bigg[1+\sum_{k=1}^{\infty}(x^{kN}\hat\I_{(k,0)}+y^{kN}\I_{(0,k)})\bigg], \qquad 
\I_{\infty} = \prod_{m=1}^{\infty}\frac{1}{(1-x^{m})(1-y^{m})}, 
\ea
with the explicit brane indices
\ba
\la{A.4}
\hat\I_{(k,0)}(x,y) &= \frac{\prod_{m=1}^{\infty}(1-y^{m})}{\prod_{m=1}^{k}(1-x^{-m})\prod_{m=1}^{\infty}(1-x^{-k}y^{m})}, \qquad
\hat\I_{(0,k)}(x,y) = \hat\I_{(k,0)}(y,x).
\ea
Taking the scaling limit (\ref{3.21}) and expanding in $\eps$ we obtain the series expansions of the finite $N$ index, for instance
\ba
\I_{1}(\eps x, \eps y) =& 1+(x+y) \eps +(x^2+y^2) \eps ^2+(x^3+y^3) \eps ^3+(x^4+y^4) \eps ^4+\cdots, \\ 
\I_{2}(\eps x, \eps y) =& 1+(x+y) \eps +(2 x^2+x y+2 y^2) \eps ^2+(x+y) (2 x^2-x y+2 y^2) \eps 
^3\lp
+(3 x^4+x^3 y+x^2 y^2+x y^3+3 y^4) \eps ^4+\cdots, \\
\I_{3}(\eps x, \eps y) =& 1+(x+y) \eps +(2 x^2+x y+2 y^2) \eps ^2+(x+y) (3 x^2-x y+3 y^2) \eps 
^3\lp
+(4 x^4+2 x^3 y+3 x^2 y^2+2 x y^3+4 y^4) \eps ^4+\cdots, 
\ea
and the expansion of the brane indices
\ba
\la{A.8}
\hat\I_{(1,0)}(\eps x, \eps y) =& 
-\frac{x^2 \eps }{x-y}-\frac{x (x^2-x y+y^2) \eps 
^2}{x-y}-\frac{(x^4-x^3 y+y^4) \eps ^3}{x-y}+\cdots, \\
\la{A.9}
\hat\I_{(2,0)}(\eps x, \eps y) =& -\frac{x^7 \eps ^4}{(x-y) y (x+y)}-\frac{x^5 (x^3-x y^2+y^3) \eps 
^5}{(x-y) y^2}-\frac{x^3 (x^8+x^7 y+x^6 y^2-x^5 y^3+x^3 y^5+y^8) \eps 
^6}{(x-y) y^3 (x+y)}+\cdots, \\
\hat\I_{(3,0)}(\eps x, \eps y) =& -\frac{x^{15} \eps ^9}{(x-y) y^3 (x^2+x y+y^2)}-\frac{x^{12} (x^6+x^4 
y^2-x^3 y^3+y^6) \eps ^{10}}{(x-y) y^5 (x^2+x y+y^2)}\lp
-\frac{x^9 
(x^{12}+x^{10} y^2+2 x^8 y^4-x^7 y^5+x^4 y^8+y^{12}) \eps 
^{11}}{(x-y) y^7 (x^2+x y+y^2)}+\cdots.
\ea
In this case, the structure of the expansion is more complicated than for the Schur index. Indeed, all terms of $\hat\I_{(k,0)}$ are non-trivial rational functions
of the fugacities. Of course, the wall-crossing poles cancel to reproduce the gauge theory index. Going to the wall $x=y$ (we omit a discussion of the other special 
points like $x=-y$) one finds that the level $k$ combination 
\be
L_{k} = x^{kN}\hat\I_{(k,0)}(x,y)+y^{kN}\hat\I_{(0,k)}(x,y) ,
\ee
has limit
\ba
\la{B12}
L_{1}|_{x=y} &= -x^{N}[(N+2)\,x+(N+1)\,x^{2}+(N-2)\,x^{3}+(N-5)\,x^{4}+\cdots], \notag  \\
L_{2} |_{x=y}&= -x^{2N}[(N+4)\,x^{4}+2(N+4)\,x^{5}+(4N+15)\,x^{6}+2(3N+10)\,x^{7}+\cdots], \notag \\
L_{3} |_{x=y}&= -x^{3N}[(N+6)\,x^{9}+2(N+6)\,x^{10}+5(N+6)\,x^{11}+(9N+53)\,x^{12}+\cdots],
\ea
These finite size corrections reproduce the giant graviton-type expansion of the index that follows from  (\ref{B.2}). Indeed,
integrating around $\zeta=0$ we have 
\be
\la{A13}
\I_{N}(x,x) = \oint\frac{d\zeta}{2\pi i }\frac{1-\zeta}{\zeta^{N+1}}\prod_{n=0}^{\infty}\frac{1}{(1-\zeta x^{n})^{2}}.
\ee
Deforming the contour to encircle the poles at $\zeta=x^{-k}$ we can identify $L_{k}$ at $x=y$ with the opposite of the residue of 
(\ref{A13}) at $\zeta=x^{-n}$
\ba
L_{k}|_{x=y} &= -\Res_{k} = -x^{-2k}
\lim_{\zeta\to x^{-k}}\frac{d}{d\zeta}\bigg[\frac{1-\zeta}{\zeta^{N+1}}\mathop{\prod_{n=0}^{\infty}}_{n\neq k}\frac{1}{(1-\zeta x^{n})^{2}}\bigg].
\ea
In particular the leading term at small $x$ is 
\ba
L_{k}|_{x=y} =&
 x^{k(N-1)}((N+1)x^{k}-N)\mathop{\prod_{n=0}^{\infty}}_{n\neq k}\frac{1}{(1- x^{n-k})^{2}}-2x^{k(N-2)}(x^{k}-1)
\mathop{\prod_{n=0}^{\infty}}_{n\neq k}\frac{1}{(1- x^{n-k})^{2}}\mathop{\sum_{n=0}^\infty}_{n\neq k}
\frac{x^{n-k}}{1- x^{n-k}} \lp
= \mathop{\prod_{n=0}^{\infty}}_{n\neq k}\frac{1}{(1- x^{n-k})^{2}}\bigg[x^{k(N-1)}((N+1)x^{k}-N)-2x^{k(N-2)}(x^{k}-1)\mathop{\sum_{n=0}^\infty}_{n\neq k}
\frac{x^{n-k}}{1- x^{n-k}} \bigg] \lp
 -\prod_{n=0}^{k-1}\frac{1}{(1- x^{n-k})^{2}}x^{k(N-1)}(N-2k+\mc O(x))  = -x^{kN+k^{2}}(N-2k+\mc O(x)),
\ea
in agreement with the first term in (\ref{B12}). Besides, the above calculation shows that higher order powers of $x$ take the form 
\be
L_{k}|_{x=y} = x^{kN+k^{2}}\sum_{p=0}^{\infty}(a_{k,p}N+b_{k,p})\,x^{p},
\ee
for some numerical coefficients $a_{k,p}$, $b_{k,p}$, again in agreement with (\ref{B12}).
 The fact that this case the enhancement of degeneracy is just linear in $N$ is
 a consequence of the fact that the singularity at $x=y$ in the brane indices is only of the simple form $1/(x-y)$ with no powers.
This is due to the $m=k$ factor in the $x,y$ dependent infinite product in (\ref{A.4}).

\bibliography{BT-Biblio}
\bibliographystyle{JHEP-v2.9}
\end{document}